\documentclass[aps,prb,twocolumn, titlepage,showpacs]{revtex4}

\usepackage{graphicx}
\usepackage{color}
\usepackage{dcolumn}
\usepackage{amsfonts}
\usepackage{bm}

\usepackage{epstopdf}

\begin{document}

\title{Oscillation-like diffusion of two-dimensional liquid dusty plasmas on one-dimensional periodic substrates with varied widths}

\author{W. Li$^1$, C. Reichhardt$^2$, C. J. O. Reichhardt$^2$, M. S. Murillo$^3$, and Yan Feng$^1,^4,^\ast$}
\affiliation{
$^1$ Center for Soft Condensed Matter Physics and Interdisciplinary Research, School of Physical Science and Technology, Soochow University, Suzhou 215006, China\\
$^2$ Theoretical Division, Los Alamos National Laboratory, Los Alamos, New Mexico 87545, USA\\
$^3$ Department of Computational Mathematics, Science and Engineering, Michigan State University, East Lansing, Michigan 48824, USA\\
$^4$ National Laboratory of Solid State Microstructures, Nanjing University, Nanjing 210093, China\\
$\ast$ E-mail: fengyan@suda.edu.cn}

\date{\today}

\begin{abstract}

The long-time diffusion of two-dimensional dusty plasmas on a one-dimensional periodic substrate with varied widths 
is investigated using Langevin dynamical simulations.
When the substrate is narrow and the dust particles form a single row, the diffusion is the smallest in both directions. We find that as the substrate width gradually increases to twice its initial value, the long-time diffusion of the two-dimensional dusty plasmas first increases, then decreases, and finally increases again, giving an oscillation-like diffusion with varied substrate width. 
When the width increases to a specific value, the dust particles within each potential well arrange themselves in a stable zigzag pattern, greatly reducing the diffusion, and leading to the observed oscillation in the diffusion with the increasing width.
In addition, the long-time oscillation-like diffusion is consistent with the number of dust particles that are hopping across the potential wells of the substrate.
\end{abstract}

\maketitle

\section{Introduction}

Diffusive motion on a substrate is a fundamental transport problem with numerous applications in various fields of science and technology~\cite{Risken:1989, Moss:1989, Jung:1993, Ala-Nissila:1992}. Two-dimensional (2D) diffusion is of great interest~\cite{Alder:1970,Ernst:1970,Dorfman:1970,Camp:2005}, and it has been widely investigated in many 2D systems, such as colloidal suspensions~\cite{Murray:1990}, strongly correlated electrons on the surface of liquid helium~\cite{Perera:1998} and, in particular, strongly coupled dusty plasmas~\cite{Liu:2008}. 
When a periodic substrate is applied to these systems, the arrangement of particles is distorted as the particles tends to move toward the local energy minima, producing a structure that is determined by the substrate. The diffusive motion of particles on the substrate is also of interest, and this is the focus of the present work.

A dusty plasma~\cite{Shukla:2002,Fortov:2005,Morfill:2009,Piel:2010,Bonitz:2010, Feng:2008}, or a complex plasma, refers to a partially ionized gas containing micron-sized particles of solid matter, called dust particles. 
Under laboratory conditions, these dust particles are typically charged to $\approx -10^4e$, and their mutual repulsion can be described by a Yukawa or Debye-H\"uckel potential,  $\phi(r)=Q^2\exp(-r/\lambda_D)/4\pi\epsilon_0r$, produced by the shielding effects of free electrons and ions in plasmas~\cite{Yukawa:1935,Melzer:1999,Konopka:2000}. Here $Q$ is the particle charge and $\lambda_D$ is the Debye screening length. Due to their extremely low charge-to-mass ratio, the dust particles are in the strongly coupled limit,
causing them to exhibit typical liquid-~\cite{Konopka:2000,Donko:2004} or solid-like~\cite{Feng:2011,Qiao:2003} properties.
In typical laboratory conditions, these charged dust particles can self-organize into a single layer, i.e., a 2D suspension~\cite{Feng1:2008,Feng1:2011} with negligible out-of-plane motion.
Dusty plasmas are amenable to direct video imaging and individual particle tracking, allowing the motion of individual dust particles to be investigated at the kinetic level~\cite{Morfill:2009, Feng1:2008, Liu2:2008}.  The dusty plasma is recognized as a promising model system in which to investigate many physical processes in solids and liquids, such as heat conduction~\cite{Nunomura:2005,Nosenko:2008,Feng:2012,Feng2:2012}, shear viscosity~\cite{Nosenko:2014} and diffusion~\cite{Juan:1998,Quinn:2002,Liu:2008,Liu:2006,Liu:2007,Ott:2008,Donko:2009,Nunomura:2006,Feng:2014}. Previous diffusion studies in 2D dusty plasmas (2DDP)~\cite{Juan:1998,Quinn:2002,Liu:2008,Liu:2006,Liu:2007,Ott:2008,Donko:2009,Nunomura:2006,Feng:2014} focused on systems without a substrate, while the diffusion of 2DDP on a substrate was briefly considered in Ref.~\cite{Wang:2018}.
To our knowledge,
diffusion of 2DDP on a one-dimensional periodic substrate (1DPS) with varied width
has not been studied previously.

Here, we report our systematic investigation of the diffusive motion of 2DDP on 1DPS with varied widths. We perform Langevin dynamical simulations to mimic a 2D dusty plasma liquid on 1DPS~\cite{Feng:2011,Li:2018,Schweigert:2000,Donko:2010}. We find that, as the width of 1DPS varies, the long-time diffusion of 2DDP exhibits an oscillation-like behavior, which we attribute to the stable structure of dust particles within the potential wells of 1DPS combined with hopping~\cite{Asaklil:2003,Porto:2001,Ferrando:2000} of dust particles between potential wells. This paper is organized as follows. In Sec.~II, we briefly explain the Langevin dynamical simulation method used here. In Sec.~III, we present our finding of the oscillation-like diffusion of 2DDP on the 1DPS with varied widths, as well as an analysis of this behavior using structural and dynamical properties. Finally, we provide a brief summary.

\begin{figure}
    \centering
    \includegraphics{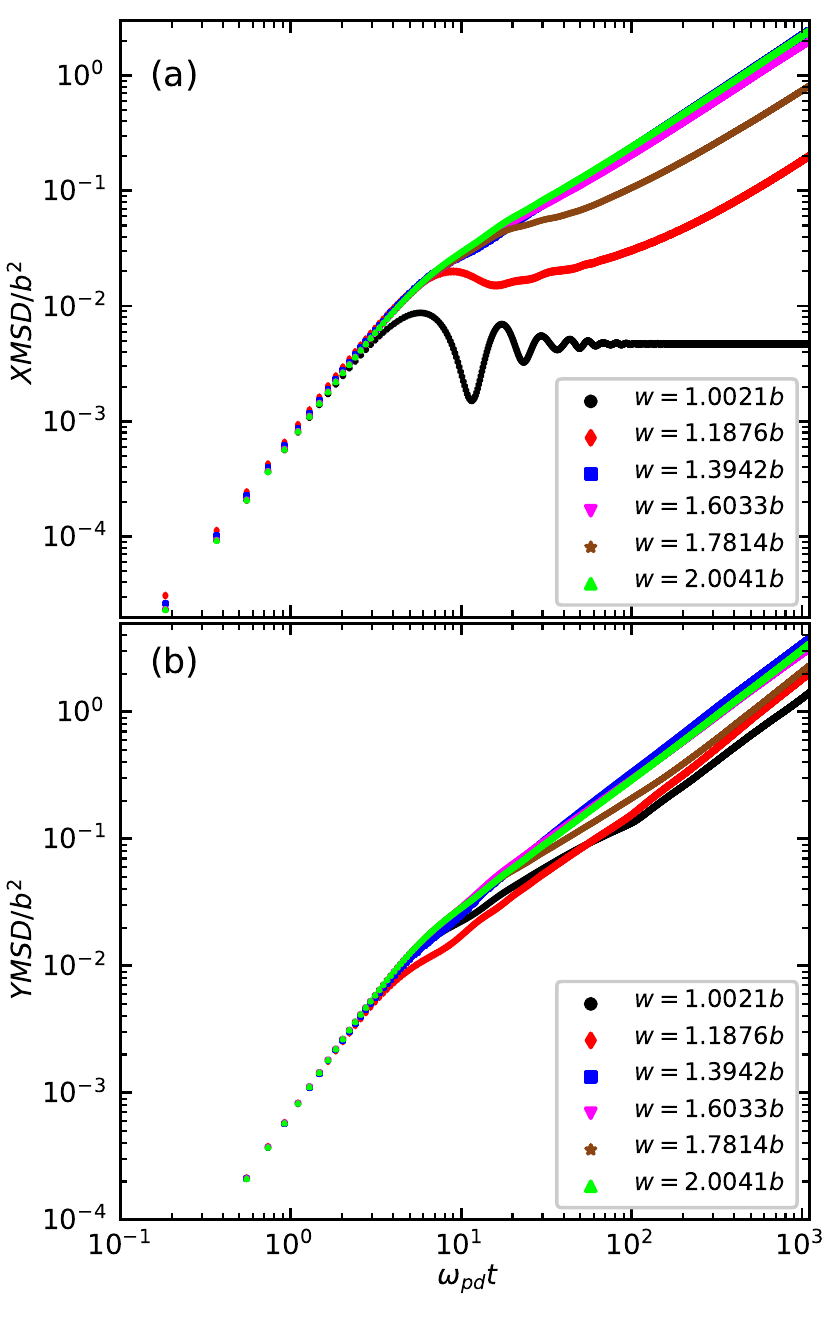}
    \caption{\label{fig:XYMSD} (Color online) Mean-squared displacement (MSD) calculated from the motion in the $x$ direction (XMSD) (a) and in the $y$ direction (YMSD) (b) for the motion of the simulated 2DDP on the 1DPS of $U_0=0.05E_0$ with the different widths. The motion in the $x$ direction is clearly strongly suppressed by the 1DPS. Also, changing the width of the 1DPS substantially modifies the XMSD and YMSD of 2DDP. We find that as the width of the 1DPS gradually increases from $1.0021b$ to $2.0041b$, the long-time diffusion (at $\omega_{pd} t =1000$) in the $x$ direction does not vary monotonically. Similarly, the long-time diffusion in the $y$ direction also exhibits multiple increases and decreases as the width of the 1DPS changes.
    }
\end{figure}

\section{Simulation method}

To study the diffusion mechanism of two-dimensional dusty plasmas on one-dimensional periodic substrates, we perform Langevin dynamical simulations of 2D Yukawa systems~\cite{Li:2018}. We numerically integrate the equation of motion
\begin{equation}\label{LDE}
{	m \ddot{\bf r}_i = -\nabla \Sigma \phi_{ij} - \nu m\dot{\bf r}_i + \xi_i(t)+{\bf F}^{S}_i,}
\end{equation}
for 1024 dust particles, confined in a rectangular box with dimensions $61.1 a \times 52.9 a$, where $a = (n\pi)^{-1}$ is the Wigner-Seitz radius for an areal number density $n$ for 2D systems. The four forces on the RHS of Eq.~(\ref{LDE}) are fully described in~\cite{Li:2018}. To characterize our 2DDP, we fix the coupling parameter $\Gamma =  Q^2/(4 \pi \epsilon_0 a k_B T) = 200$ and the screening parameter $\kappa = a/\lambda_{D} = 2$,
where $T$ is the kinetic temperature of the simulated dust particles.
In addition to the value of $a$, we use the lattice constant $b$ to normalize the length,
with $b = 1.9046a$ for a 2D defect-free triangular lattice.

The force ${\bf F}^{S}_i$ from the 1DPS is 
\begin{equation}\label{FS}
{	{\bf F}^{S}_i = - \frac {\partial U(x)}{\partial x} = (2\pi U_0/w) \sin (2\pi x/w) \hat{\bf x}, }
\end{equation}
where $U(x) = U_0 \cos(2\pi x/w)$ is an array of potential wells parallel to the $y$ axis. Here, $U_0$ is the substrate strength in units of $E_0 = Q^2/4\pi\epsilon_0 a $ and $w$ is the width of the substrate in units of $b$. We specify the substrate strength $U_0 = 0.05E_0$, and gradually change  the substrate width $w$ from $1.0021b$ to $2.0041b$. Since the simulated size is $61.1a \approx 32.07b$ in the $x$ direction, to satisfy the periodic boundary conditions, we choose $w/b = 1.0021$, $1.1057$, $1.1876$, $1.2333$, $1.2826$, $1.3942$, $1.5270$, $1.6033$, $1.6877$, $1.7814$, $1.8862$, and $2.0041$, corresponding to $32, 29, 27, 26, 25, 23, 21, 20, 19, 18, 17$, and 16 full potential wells, respectively. 

For each simulation run, we begin with a random configuration of dust particles and integrate for $3 \times 10^5$ simulation time steps to achieve a steady state. We then record the particle positions and velocities during the next $10^7$ simulation steps.  The time step is $0.0037 {\omega}_{pd}^{-1}$ and ${\omega}_{pd} = (Q^2/2\pi\epsilon_0 m a^3)^{1/2}$ is the nominal dusty plasma frequency. Other simulation details are the same as those in~\cite{Li:2018, Wang:2018}. In addition, we also perform a few test runs with a much larger system containing $4096$ dust particles, and find that there is no substantial difference in the results reported here.

\section{Results and Discussion}

\subsection{Diffusion of 2DDP on 1DPS}

Here, we study the diffusive motion of 2DDP on 1DPS with different widths. 
We calculate the time series of the mean-square displacement (MSD) from the motion of dust particles, defined as
\begin{equation}\label{MSD}
{{\rm MSD} = \langle |{\bf r}_{i}(t) - {\bf r}_{i}(0)|^{2} \rangle = 4Dt^{\alpha},}
\end{equation}
where $\langle ~ \rangle$ denotes the ensemble average and ${\bf r}_{i}(t)$ is the position of the $i$th particle at time $t$. For the long-time diffusive motion, the exponent $\alpha$ reflects the diffusion properties. The exponent $\alpha=1$ indicates normal diffusion, while $\alpha < 1$ and $\alpha > 1$ correspond to sub- and super-diffusion, respectively. 

Figure~1 shows the calculated MSD of 2DDP due to the motion in two directions on 1DPS with different widths. 
In the presence of the 1DPS, the initial ballistic and final long-time diffusive motion is accompanied by sub-diffusive motion at intermediate timescales, as clearly shown in Fig.~\ref{fig:XYMSD}(b) and first observed in~\cite{Wang:2018}. Here, we focus primarily on the long-time diffusive motion. In Fig.~\ref{fig:XYMSD}(a), we shown XMSD, which is the MSD calculated from the motion in only the $x$ direction, perpendicular to the potential wells.
Similarly, in Fig.~\ref{fig:XYMSD}(b), YMSD is calculated from the motion in only the $y$ direction, parallel to the potential wells.

\begin{figure}
	\centering
	\includegraphics{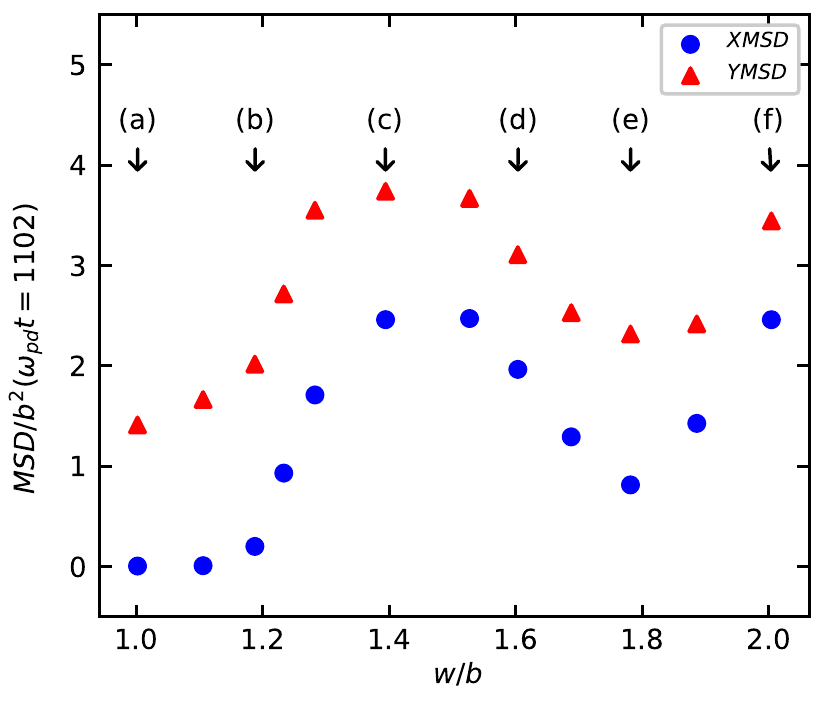}
	\caption{\label{fig:XYLMSD}(Color online) The long-time MSD at $\omega_{pd}t=1102$, determined from the motion in two directions for 2DDP on 1DPS with different widths. As the substrate width increases monotonically from $w = 1.0021b$ to $w = 2.0041b$, XMSD and YMSD both initially increase, then decrease, and finally increase again. Although XMSD is much smaller than YMSD due to the suppressed motion of dust particles in the $x$ direction by the 1DPS,  the magnitude of YMSD is exactly synchronized with that of XMSD. We speculate that there is a coupling of the motion in one direction with the motion in the other direction. To further analyze this oscillation-like long-time diffusion as the substrate width varies, we select six typical data points, which are marked with arrows.
        }
\end{figure}

As the width of the 1DPS gradually increases from $w=1.0021b$ to $2.0041b$, the long-time XMSD and YMSD (at $\omega_{pd}t=1000$) do not vary monotonically, as shown in Fig.~\ref{fig:XYMSD}. For some specific values of the width of the 1DPS, such as $w = 1.7814b$, we find a remarkable decrease of the diffusion in both directions. 
Although there is no constraint from the potential wells along $y$ direction, the long-time YMSD also varies nonmonotonically in the same way as XMSD as the 1DPS width increases. We speculate that this is due to the coupling of the  motion of dust particles in one direction with the motion in the other direction. Note that when the width of 1DPS is close to the lattice constant $b$, i.e., $w=1.0021b$, the diffusion at long timescales is completely suppressed, suggesting that almost no dust particles can hop~\cite{Asaklil:2003,Porto:2001,Ferrando:2000} out of the potential wells in which they were originally located.

To study the variation of the long-time MSD with the increasing width of the 1DPS, we plot the long-time MSD at $\omega_{pd}t=1102$ for different widths of 1DPS ranging from $w=1.0021b$ to $2.0041b$ in Fig.~\ref{fig:XYLMSD}. 
Clearly, the long-time XMSD and YMSD do not vary monotonically at all. We find that the long-time YMSD is always much larger than the long-time XMSD due to the suppressed motion of dust particles in the $x$ direction by the 1DPS.

\begin{figure}
	\centering
	\includegraphics{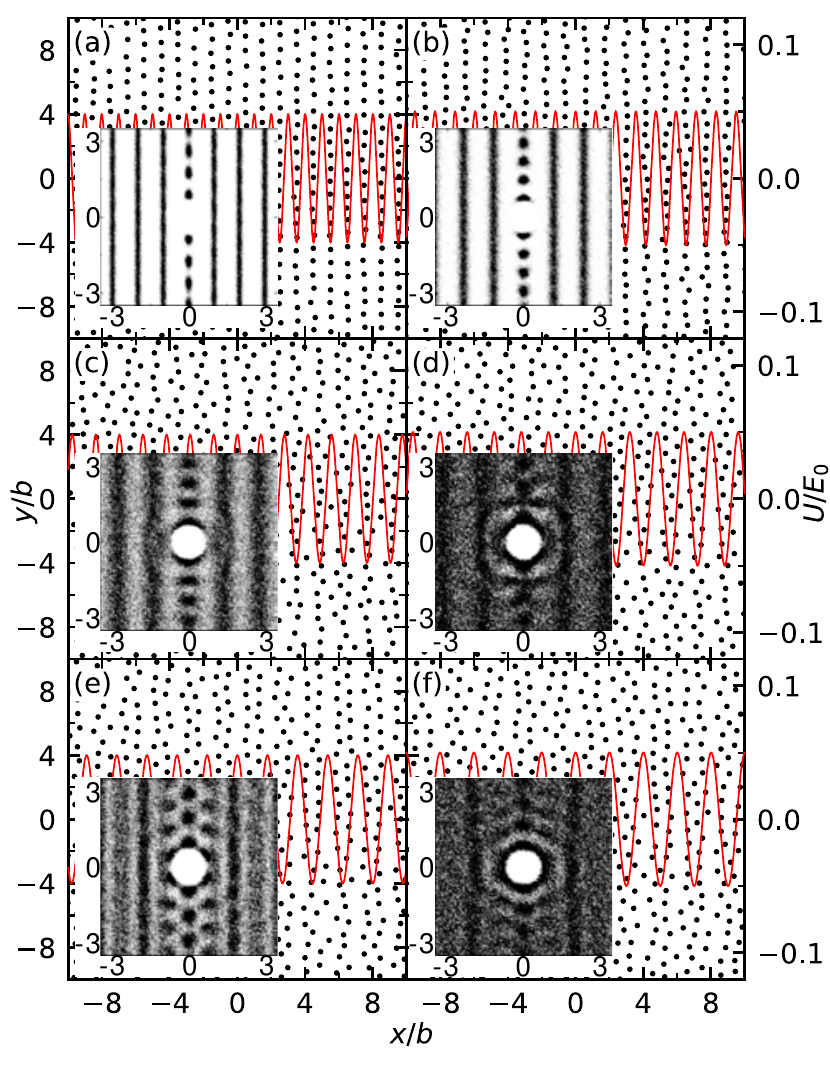}
	\caption{\label{fig:Distribution}
     (Color online).
          Snapshots of the simulated dust particle positions (dots) in a 2DDP with $\Gamma=200$ and $\kappa=2$ on the 1DPS (curves) for the six different widths indicated in Fig.~\ref{fig:XYLMSD}: $w=$ (a) $1.0021b$, (b) $1.1876b$, (c) $1.3942b$, (d) $1.6033b$, (e) $1.7814b$, and (f) $2.0041b$.  The substrate strength is fixed at $U_0 = 0.05E_0$. For each panel, the inset illustrates the corresponding 2D distribution function~\cite{Loudiyi:1992} $g(x,y)$ calculated from the simulated dust particle positions. When $w$ is small as in (a) and (b), all of the dust particles are pinned at the bottom of each potential well, forming 1D or quasi-1D chains. When the width is larger as in (c) and (d), the dust particle arrangement is much more disordered, giving a liquid-like signature in $g(x,y)$. When the width increases to $w=1.7814b$ as in (e), the dust particles form an ordered zigzag lattice arrangement within each potential well, as also clearly indicated by the features in $g(x,y)$. When the width increases further to $w=2.0041b$ as in (f), the dust particles become disordered again, giving a nearly liquid-like signature in $g(x,y)$. 
        }
\end{figure}

As the major result of this paper, we find the oscillation-like diffusion of dust particles as a function of the width of the 1DPS. As the width of 1DPS increases monotonically from $w=1.0021b$ to $w=2.0041b$, the XMSD and YMSD both initially increase, then decrease, and finally increase again, as shown in Fig.~\ref{fig:XYLMSD}. Changes in the width of the 1DPS modifies XMSD and YMSD of the 2DDP substantially. For some specific values of the width of the 1DPS, such as $w=1.7814b$, we find a remarkable decrease of the long-time diffusion.

\begin{figure}
    \centering
    \includegraphics[width=3.3in]{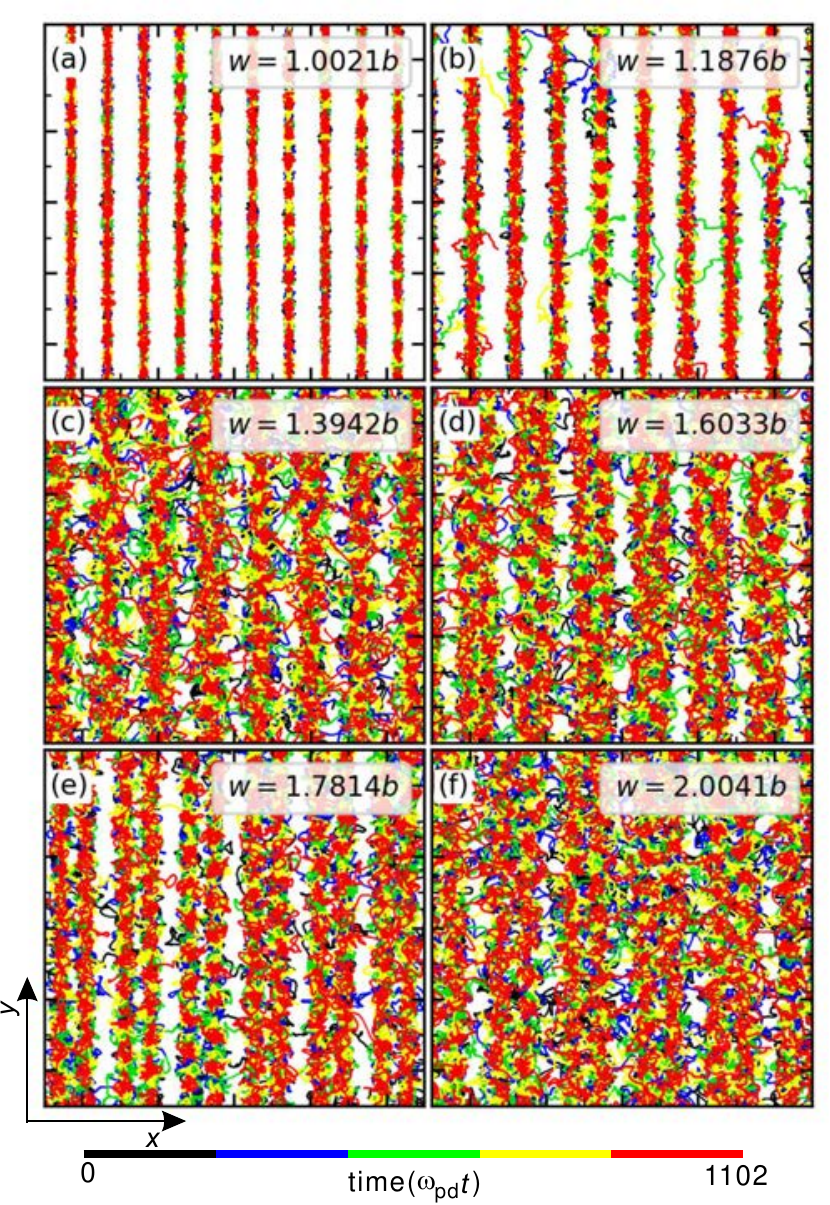}
    \caption{\label{fig:Tra}
    (Color online).
Typical trajectories of dust particles from the simulated 2DDP on the 1DPS with six different widths, for the same conditions as in Fig.~\ref{fig:Distribution}. When $w = 1.0021b$ in (a), all of the dust particles are pinned at the bottom of the potential wells, and almost no particles can hop across the potential wells of the 1DPS. As the width gradually increases to $w = 1.1876b$ in (b) and $1.3942b$ in (c), dust particles are able to move much more freely in the $x$ direction inside the potential wells, and at the same time, more and more dust particles are able to hop across the potential wells. When $w=1.6033b$ in (d), the number of hopping dust particles decreases compared to (c). When the width increases further to $w = 1.7814b$ in (e), the hopping of the dust particles greatly diminishes, and two ordered rows of dust particles can be clearly observed at the bottom of each potential well. When the width further increases to $w = 2.0041b$ in (f), the number of hopping particles increase again and the system is much more disordered.  These panels show that as the width increases from $w = 1.0021b$ to $w=2.0041b$, the number of hopping dust particles across the potential wells first increases, then decreases, and finally increases again.
     }
\end{figure}

Interestingly, the long-time YMSD is exactly synchronized with the XMSD, as indicated by the fact that XMSD and YMSD have the same oscillation behavior. We speculate that this synchronization is a result of the coupling of the motion of the dust particles in one direction with the motion in the other direction. To further investigate the underlying physics of this oscillation-like diffusion, we select six typical data points (a,b,c,d,e,f), marked with arrows in Fig.~\ref{fig:XYLMSD}, at which to analyze the static arrangement and dynamics of the dust particles.

\subsection{Structure of 2DDP on 1DPS}

To determine the underlying physics of the observed oscillation-like diffusion, we first study the structure or arrangement of the dust particles for the six different widths highlighted in Fig.~\ref{fig:XYLMSD}. 
In Figure~\ref{fig:Distribution}, we present snapshots of the simulated dust particles of the 2DDP with $\Gamma=200$ and $\kappa=2$ on 1DPS with different values of $w$ ranging from $w=1.0021b$ to $w=2.0041b$.
The inset of each panel indicates the corresponding 2D distribution function~\cite{Loudiyi:1992} $g(x,y)$ calculated from the simulated dust particle positions. The 2D distribution function $g(x,y)$ gives the probability density of finding a particle at position ${\bf r}_2$, given that a particle is located at ${\bf r}_1$. Unlike the pair correlation function $g(r)$ widely used for isotropic systems, the 2D distribution function $g(x,y)$ is the static structural measure employed for anisotropic systems such as ours. Using
$g(x,y)$, we can clearly distinguish whether the structure of the dust particles is ordered or disordered.

As the width of 1DPS gradually increases from $w=1.0021b$ to $2.0041b$, the structure of the 2DDP within the 1DPS first changes from ordered to disordered, then orders again, and finally returns to a disordered state, as shown in Fig.~\ref{fig:Distribution}. Initially when $w$ is small, as shown in Figs.~\ref{fig:Distribution}(a) and \ref{fig:Distribution}(b), all of the dust particles are pinned at the bottom of each potential well, forming 1D or quasi-1D chains. The constraint from the substrate is larger when the width of the 1DPS is smaller, so the dust particles are in an ordered arrangement for small $w$ and their long-time diffusive motion is greatly suppressed in both the $x$ and $y$ directions. 
When the width is larger as in Figs.~\ref{fig:Distribution}(c) and \ref{fig:Distribution}(d), the dust particle arrangement is much more disordered as clearly observed from the liquid-like distribution function $g(x,y)$, and at the same time the long-time diffusive motion of the dust particles is much larger. 
When the width of 1DPS increases to $w=1.7814b$ as shown in Fig.~\ref{fig:Distribution}(e), we find that, within each potential well, the dust particles form a stable ordered zig-zag arrangement.  From the calculated 2D distribution function $g(x,y)$ in Fig.~\ref{fig:Distribution}(e), this zigzag structure produces sixfold-symmetric enhanced peaks in probability around the center. The long-time diffusive motion of the dust particles is greatly suppressed by this stable zigzag structure for the 1DPS of width $w=1.7814b$. When the width of the 1DPS increases further to $w=2.0041b$ as shown in Fig.~\ref{fig:Distribution}(f), the dust particles become disordered again, forming a nearly liquid-like state as indicated by the ring-like signature in $g(x,y)$. As a result, the diffusive motion of the dust particles increases again. 

The changing trend of the structure of the 2DDP in Fig.~\ref{fig:Distribution} matches well with the oscillation-like diffusion for varied widths of the 1DPS. The arrangement of the dust particle at the specific width of $w=1.7814b$ results in a dramatic decrease in  the long-time diffusion due to the stable zigzag structure of the 2DDP. We speculate that the dynamics of the 2DDP, such as the hopping of dust particles across the potential wells of the 1DPS, would also be substantially affected by this stable structure, as we study below. 

\subsection{Dynamical behavior of 2DDP on 1DPS}

To verify our speculation regarding the hopping of dust particles across the potential wells of the 1DPS, we measure typical trajectories over a time duration of $\omega_{pd}t=1102$ for the dust particles from the simulated 2DDP on 1DPS with six different widths, as shown in Fig.~\ref{fig:Tra}, with the same conditions as in Fig.~\ref{fig:Distribution}.

When $w=1.0021b$ in Fig.~\ref{fig:Tra}(a), all of the particles are pinned at the bottom of the potential wells, and almost no particles can overcome the potential barrier to hop across the potential wells.
As the width gradually increases from $w=1.1876b$ in Fig.~\ref{fig:Tra}(b) to $w=1.3942b$ in Fig.~\ref{fig:Tra}(c), the dust particles are able to move much more freely in the $x$ direction inside the potential wells. More and more dust particles are able to overcome the potential barrier to hop across the potential wells. As a result, the long-time MSD is much larger for these widths. When $w=1.6033b$ in Fig.~\ref{fig:Tra}(d), within each potential well we find that two rows of dust particles are able to form due to the increased width of the 1DPS.  The number of hopping dust particles is, however, smaller than in Fig.~\ref{fig:Tra}(c), most likely due to the increased repulsion of particles in adjacent columns that results from the smaller spacing between the particles in the buckled configuration.
When the width increases to $w=1.7814b$ in Fig.~\ref{fig:Tra}(e), two rows of dust particles can be clearly observed at the bottom of each potential well, corresponding to the stable zigzag arrangement, and coinciding with a dramatic reduction of the number of hopping dust particles.
When the width increases further to $w=2.0041b$ in Fig.~\ref{fig:Tra}(f), the number of hopping dust particles increases again, which is likely facilitated by the
the disordered arrangement of the dust particles, as shown in Fig.~\ref{fig:Distribution}(f).

\begin{figure}
	\centering
	\includegraphics{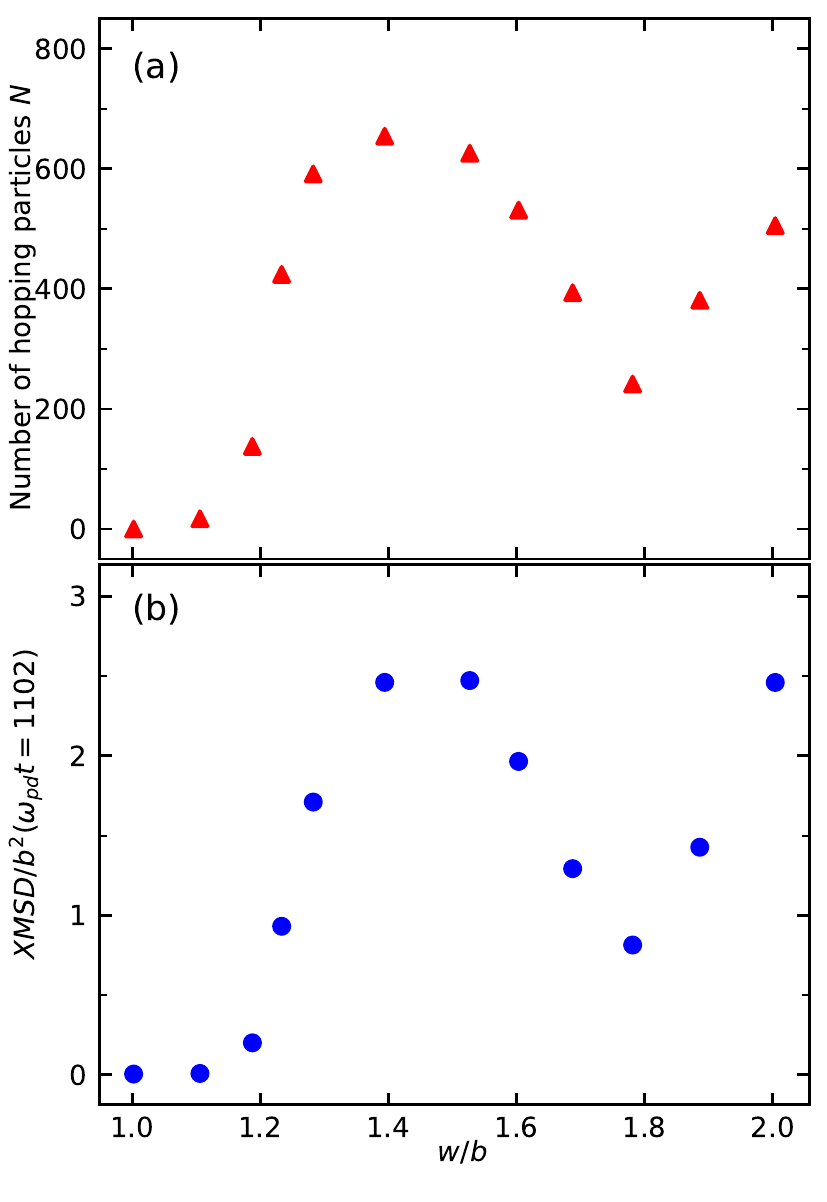}
	\caption{\label{fig:JumNum}
     (Color online).
(a) The number $N$ of dust particles hopping across the potential wells and (b) the long-time MSD due to the motion of dust particles in the $x$ direction during a time duration of $\omega_{pd}t=1102$ as functions of the width $w$ of the 1DPS for a constant strength of $U_0=0.05E_0$. Clearly, the XMSD in (b) is synchronized with the number of hopping dust particles $N$ in (a), since the hopping dust particles make a much larger contribution to XMSD. As the width of the 1DPS increases from $w=1.0021b$ to $1.3942b$, the number of hopping dust particles $N$ gradually increases, since more and more dust particles are able to move across the potential wells. When the width increases further from $w=1.3942b$ to $1.7814b$, the number of hopping dust particles gradually diminishes. We attribute this to the appearance of a stable zigzag arrangement of dust particles, as also shown in Figs.~\ref{fig:Distribution} and~\ref{fig:Tra}. When the width increases further to $w=2.0041b$, the number of hopping dust particles $N$ gradually increases again.
        }
\end{figure}

Note that, as the width of the 1DPS increases from $1.0021b$ to $2.0041b$, the spatial region inside each potential well over which the motion of dust particles in the $x$ direction occurs first increases, then diminishes, and finally increases again. At some widths, such as $w=1.7814b$ in Fig.~\ref{fig:Tra}(e), two rows of dust particles clearly appear at the bottom of each potential well, corresponding to the stable zigzag arrangement, and there is probably some diffusive motion between these two rows. However,
in the MSD measurement, and especially in XMSD, the diffusive motion between rows inside each potential well
has a much smaller effect than the hopping diffusion between potential wells, since the displacement due to inter-well hopping is much larger.

We next quantify the hopping dust particles by
counting the number $N$ of hopping dust particles from our simulation data, as shown in Fig.~\ref{fig:JumNum}(a). For comparison, we also plot the corresponding XMSD in Fig.~\ref{fig:JumNum}(b). Clearly, the long-time XMSD in Fig.~\ref{fig:JumNum}(b) due to the motion of dust particles in the $x$ direction is synchronized with $N$
as a function of the width of 1DPS for the time duration of $\omega_{pd}t=1102$. As we speculated above, the long-time XMSD is determined mainly by the number $N$ of dust particles that hop across the potential wells, and $N$ is affected by the structural arrangement of the dust particles inside the potential wells.

As the width of the 1DPS increases from $w=1.0021b$ to $2.0041b$, the number of hopping dust particles $N$ first increases, then decreases, and finally increases again. When the width of 1DPS increases from $w=1.0021b$ to $1.3942b$, the number of hopping dust particles $N$ gradually increases since more and more dust particles can overcome the potential energy barrier to hop across the potential wells due to the unstable disordered arrangement of dust particles inside the potential wells, as shown in Figs.~\ref{fig:Distribution} and~\ref{fig:Tra}. When the width increases further from $w=1.3942b$ to $1.7814b$, the number of hopping dust particles $N$ gradually decreases due to the emergence of the stable zigzag arrangement of dust particles inside the potential wells, as also shown in Figs.~\ref{fig:Distribution} and~\ref{fig:Tra}. When the width increases further to $w=2.0041b$, the number of hopping dust particles $N$ gradually increases again, since now the arrangement of dust particles changes from zigzag to a nearly liquid-like state. Thus, as the width of the 1DPS varies, the oscillation-like diffusion is correlated with the static arrangement of the dust particles which modifies the dynamical dust particle hopping across the potential wells of the 1DPS.

\section{Summary}
In summary, using Langevin dynamical simulations, we study the diffusion of 2DDP on the 1DPS. We find that, as the width of the 1DPS increases, the diffusion of 2DDP exhibits an oscillation-like feature. For small 1DPS widths, the dust particles arrange into a single row at the bottom of each potential well and the diffusion is greatly suppressed. As the width of the 1DPS gradually increases, the diffusion of the dust particles first increases, then decrease, and finally increases again. Based on the static structural measures and trajectories, we attribute the diffusion decrease to the stable zigzag arrangement of dust particles within the potential well that occurs when the width is about $w=1.7814b$. We also find that the long-time diffusion of 2DDP in one direction is synchronized with the diffusion in the other direction.
Future directions include studying the sliding dynamics under a drive or tilt to see whether
different types of depinning correlate with the oscillations in the diffusion or whether different types of creep behavior occur \cite{review}.

\section{Acknowledgments}
Work in China was supported by the National Natural Science Foundation of China under Grants No. 11875199 and No. 11505124, the 1000 Youth Talents Plan, startup funds from Soochow University, and the Priority Academic Program Development (PAPD) of Jiangsu Higher Education Institutions. Work at LANL
was supported by the US Department of Energy through
the Los Alamos National Laboratory.  Los Alamos National Laboratory is
operated by Triad National Security, LLC, for the National Nuclear Security
Administration of the U. S. Department of Energy (Contract No. 892333218NCA000001).

\end{document}